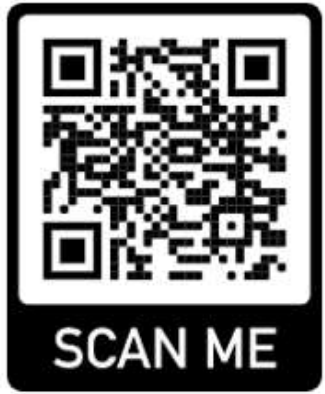

# Advanced Drone Swarm Security by Using Blockchain Governance Game

SONG-KYOO (AMANG) KIM

**ABSTRACT**

This research contributes to the security design of an advanced smart drone swarm network based on a variant of the *Blockchain Governance Game* (BGG), which is the theoretical game model to predict the moments of security actions before attacks, and the *Strategic Alliance for Blockchain Governance Game* (SABGG), which is one of the BGG variants which has been adapted to construct the best strategies to take preliminary actions based on strategic alliance for protecting smart drones in a blockchain-based swarm network. Smart drones are artificial intelligence (AI)-enabled drones which are capable of being operated autonomously without having any command center. Analytically tractable solutions from the SABGG allow us to estimate the moments of taking preliminary actions by delivering the optimal accountability of drones for preventing attacks. This advanced secured swarm network within AI-enabled drones is designed by adapting the SABGG model. This research helps users to develop a new network-architecture-level security of a smart drone swarm which is based on a decentralized network.

**Keywords:** Drone, swarm, Blockchain Governance Game; artificial inelligence, mixed game; stochastic model; fluctuation theory; 51 percent attack

## I. INTRODUCTION

Drones occupy an essential place in both military and civilian applications for various roles including criminal investigations, public safety organizations, transportation management facilities, and surveillance forces [1]. Because of dynamic mobility, quick reaction and easy deployment, drones offer new possibilities for different applications with affordable expense [2]. A drone swarm is multiple drones being used at once and drones in a swarm communicate and collaborate, making collective decisions of collective actions. In a militarized drone swarm, instead of 10 or 100 distinct drones, the swarm forms a single, integrated weapon system guided by some form of artificial intelligence [3]. The Blockchain Governance Game (BGG) has been designed as a stochastic game model with the fluctuation and the mixed strategy game for analyzing the network to provide the decision making moment for taking preliminary security actions before attacks. The model is targeted to prevent blockchain based attacks (i.e., the 51 percent attack) and keeps the network decentralized. Atypical case which an attacker tries to build an alternative blockchain (blockchain forks) faster than regular miners [18].

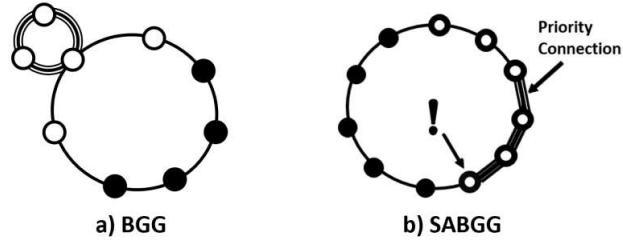

**Fig 2.** BGG vs. SABGG [18, 21]

## II. STOCHASTIC GAME FOR ASDS SECURITY FRAMEWORK

The proposed ASDS network structure is considered and the drones in a swarm are connected each other and a swarm is hooked up as single Blockchain network (see Fig. 2). Drones in a swarm are fully connected but these may not connected with a command center (or a control center). This drone swarm could execute thier command artificially and independently even with disconection with a command center. Each drone randomly generates unique data (e.g., GPS cordinates, motor RPM values) and broadcase these data to other drones (which is equivalant with transaction in a blockchain network). Each drone generates the value based on its mechanical action and the generated values are sharing with all other drones in a swarm.

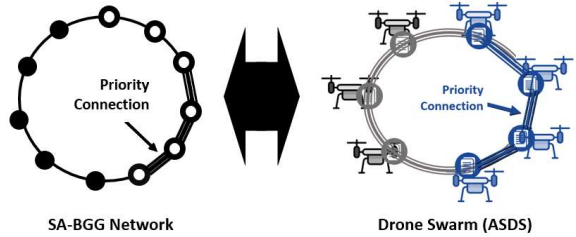

**Fig 2.** Adapting SA-BGG for the ASDS network architecture

To apply the SABGG into the ASDS network structure, the antagonistic game of two players (called "A" and "H") are introduced to describe the Blockchain network in a drone swarm as a defender and an attacker. The joint functional of the Blockchain network model with the strategic alliance is as follows:

$$\Phi_{\lceil \frac{M}{2} \rceil}(\xi, g_0, g_1, b, z_0, z_1) \tag{2.16}$$

$$= \mathbb{E}\left[\xi^\nu \cdot g_0^{A_{\nu-1}} \cdot g_1^{A_\nu} \cdot b^{A_\nu - B} \cdot z_0^{H_{\mu-1}} \cdot z_1^{H_\mu} \mathbf{1}_{\{\nu < \nu_2 < \mu\}}\right],$$

$$\|\xi\| \leq 1, \|g_0\| \leq 1, \|g_1\| \leq 1, \|b\| \leq 1, \|z_0\| \leq 1, \|z_1\| \leq 1. \tag{2.17}$$

where $M$ indicates the total number of nodes (or ledgers) in the swarm network for each drone (see Fig. 2). The Theorem of SABGG establishes an explicit formula $\Phi_{\lceil \frac{M}{2} \rceil}$ from (2.7)-(2.10). Based on the theorem [25], the functional $\Phi_{\lceil \frac{M}{2} \rceil}$ of the process of (2.16) satisfies following expression:

$$\Phi_{\lceil \frac{M}{2} \rceil}(\xi, g_0, g_1, z_0, z_1) = \mathfrak{D}_{(q,r,s)}^{(\lceil \frac{M}{2} \rceil, \lceil \frac{M}{2} \rceil, \lceil \frac{M}{2} \rceil)} \Lambda, \tag{2.18}$$

where

$$\Lambda = \sigma \cdot \Gamma \left( \frac{1 - \Gamma^1}{1 - \Gamma} \right) \left( \gamma_0^1 - \gamma_0 + \frac{\zeta \Theta_0}{1 - \zeta \Theta} (\gamma^1 - \gamma) \right), \tag{2.19}$$

and

$$\Theta := \gamma(g_0 g_1 b q r, z_0 z_1 s), \tag{2.20}$$
$$\Theta_0 := \gamma_0(g_0 g_1 b q r, z_0 z_1 s), \tag{2.21}$$

$$\gamma := \gamma(g_1 b q, z_1), \tag{2.22}$$
$$\gamma_0 := \gamma_0(g_1 b q, z_1), \tag{2.23}$$

$$\gamma^1 := \gamma(g_1 b, z_1), \tag{2.24}$$
$$\gamma_0^1 := \gamma_0(g_1 b, z_1), \tag{2.25}$$

$$\Gamma := \gamma(br, s), \tag{2.26}$$
$$\Gamma^1 := \gamma(r, 1), \tag{2.27}$$

$$\sigma := \mathbb{E}[b^{-B}]. \tag{2.28}$$

From (2.13)-(2.14), we can find the PGFs (probability generating functions) of the *exit index* $\nu$:

$$\mathbb{E}[\xi^\nu] = \Phi_{\lceil \frac{M}{2} \rceil}(\xi, 1, 1, 1, 1) \tag{2.32}$$

Let us consider a two-person mixed strategy game, and player H (i.e., a drone swarm) is the person who has two strategies at the observation moment, one step before attackers complete to generate alternative chains with dishonest transactions. In this case, the cost will be not only all drones in a swarm but also the alliance costs. The normal form of games is as follows:

.  Players: $\quad \boldsymbol{N} = \{A, H\}, \tag{2.37}$
.  Strategy sets:
$$\boldsymbol{s_a} = \{\text{"NotBurst"}, \text{"Burst"}\},$$
$$\boldsymbol{s_h} = \{\text{"Regular"}, \text{"Safety"}\}.$$

Based on the above conditions, the general cost matrix at the prior time to be burst $\tau_{\nu-1}$ could be composed as follows:

**Table 1.** Cost matrix

|          | NotBurst $(1-q(s_h))$ | Burst $(q(s_h))$ |
|----------|-----------------------|------------------|
| Regular  | 0                     | $V$              |
| Safety   | $c_b$                 | $c_b + V$        |

where $q(s_h)$ is the probability of bursting blockchain network (i.e., an attacker wins the game) and it depends on the strategic decision of player H:

$$q(s_h) = \begin{cases} \mathbb{E}\left[\mathbf{1}_{\{A_\nu \geq \frac{M}{2}\}}\right], & s_h = \{Regular\}, \\ \mathbb{E}\left[\mathbf{1}_{\{A_\nu - B \geq \frac{M}{2}\}}\right], & s_h = \{Safety\}, \end{cases} \quad (2.38)$$

and the alliance (i.e., "Safety" strategy of player H) cost should be less than the cost of other strategies. Otherwise, player H does not have to spend the cost of the strategic alliance with genuine drones. Recalling from (2.38), the probability of bursting a Blockchain network (i.e., an attacker wins the game) under the memoryless properties becomes the Poisson compound process:

$$q(s_h) = \begin{cases} \sum_{k > \frac{N}{2}} \mathbb{E}\left[\mathbf{1}_{\{A_\nu = k\}}\right], & s_h = \{Regular\}, \\ \mathbb{E}\left[\mathbb{E}\left[\sum_{k > \frac{N}{2} + B} \mathbb{E}\left[\mathbf{1}_{\{A_\nu = k\}}\right] \Big| B\right]\right], & s_h = \{Safety\}, \end{cases} \quad (2.39)$$

where

$$\mathbb{E}\left[\mathbf{1}_{\{A_\nu = k\}}\right] = \mathbb{E}\left[\mathbb{E}\left[\frac{\lambda_a \tau_\nu}{k!} \cdot e^{-\lambda_a \tau_\nu} \Big| \tau_\nu\right]\right]. \quad (2.40)$$

### III. THE ASDS OPTIMIZATION PRACTICE

A network security in an ASDS network is considered in this subsection. The strategy for protecting the ASDS is for priority connection with neighbor drones to give the less chance that an attacker catches blocks with false control requests. The example in this paper is targeting 20 drones in single swarm and each estimated drone value is around 1,500 USD in the swarm (see Table II).

| Name | Value | Description |
|------|-------|-------------|
| $M$ | 20 [Drones] | Total number of the nodes in a drone swarm |
| $V$ | 1,500 [USD/Drone] $\cdot$ $M$ [Drone] | Total value of a Blockchain enabled swarm |
| $c(\varrho)$ | $= 3\left(\frac{M}{2} - 1\right) \cdot \varrho$ [USD] | Cost for reserving nodes to avoid attacks per each car |
| $\mathbb{E}[\nu]$ | 3 [Trial] | Total number of blocks that changed by an attacker at $\tau_0 (= 0)$ |
| $B$ | $-$ | Number of accepted allys at $\tau_\nu$ |

**Table II.** Initial conditions for the cost function

Based on the above conditions, the LP (Linear Programing) model could be described as follows from (2.43)-(2.46):

Objective
 minimizing $G = \mathfrak{S}(\varrho)_{\text{Total}}$ (3.30)

Subject to

$$n \geq \frac{c(\varrho)}{V \cdot q^0 - c(\varrho)}; \tag{3.31}$$

From (2.46), the total cost $\mathfrak{S}(\varrho)_{\text{Total}}$ is as follows:

$$\mathfrak{S}(\varrho)_{\text{Total}} = \left(c(\varrho)\left(1 - q_\eta^1\right) + (c(\varrho) + V)q^1(\varrho)\right) p_{A_{-1}} \tag{3.32}$$

$$+ V \cdot q^0 (1 - p_{A_{-1}})$$

where

$$p_{A_{-1}} = \boldsymbol{P}\left\{A_{\nu-1} < \frac{M}{2}\right\}$$

$$\simeq \boldsymbol{P}\left\{A_\nu < \frac{M}{2} - \lambda_a \widetilde{\delta}\right\}$$

$$= \sum_{k=0}^{\left\{\frac{M}{2} - \lambda_a \widetilde{\delta}\right\}} \left(\frac{\{\lambda_a(\widetilde{\delta}_0 + \mathbb{E}[\nu-1]\widetilde{\delta})\}^k}{k!} \cdot e^{-\lambda_a(\widetilde{\delta}_0 + \mathbb{E}[\nu-1]\widetilde{\delta})}\right), \tag{3.33}$$

$$q^0 \simeq 1 - \sum_{k=0}^{\frac{M}{2}} \left(\frac{\{\lambda_a(\widetilde{\gamma}_0 + \mathbb{E}[\nu-1]\widetilde{\gamma})\}^k}{k!} \cdot e^{-\lambda_a(\widetilde{\gamma}_0 + \mathbb{E}[\nu-1]\widetilde{\gamma})}\right) \tag{3.34}$$

$$q^1(\varrho) = \sum_{j=0}^{\frac{M}{2}-1} \sum_{\{k \geq \frac{M}{2}+j\}} \left(\frac{\lambda_a(\widetilde{\delta}_0 + \mathbb{E}[\nu-1]\widetilde{\delta})}{k!} \cdot e^{-\lambda_a(\widetilde{\delta}_0 + \mathbb{E}[\nu-1]\widetilde{\delta})}\right) P_j, \tag{3.35}$$

$$P_j = \binom{\frac{M}{2}-1}{j} \varrho^j (1-\varrho)^{\frac{M}{2}-1-j}. \tag{3.36}$$

The total cost $\mathfrak{S}(\varrho)_{\text{Total}}$ could be minimized by given $\varrho$ is the optimal value of the reserved nodes. The below illustration in Fig. 3 is atypical graph of an optimal result by using the SABGG based ASDS network based on the given conditions in Table II.

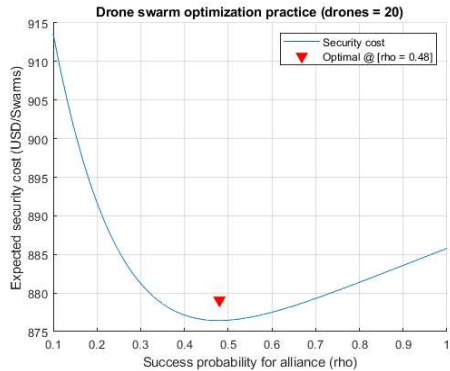

**Fig. 3.** Optimization Example for the ASDS

## IV. Conclusion

An advanced secure drone swarm network architecture protects a drone swarm from an attacker by adapting a blockchain governance game variant. The~Strategic Alliance for Blockchain Governance Game (SABGG) which is an analytically proven game model has been applied as a blockchain governance game variant. The SABGG has been adapted for a decentralized network to improve drone swarm security. The~special SABGG case demonstrates how the theoretical model is actually implemented for smart drone security. Although~this research is still theoretical and there are several steps remaining for actual implementation into real drones, the practical case demonstrates how an SABGG network could be implemented for smart drone securities and its feasibility. This paper is the first piece of research that applies an SABGG model into a swarm network architecture security. The~advanced smart drone swarm network is the successor of blockchain-governance-game-based IoT security applications, particularly in the intelligent military domain. The managerial aspects and actual implementations of smart drone operations could be the next step. Additionally, expending the domains for applying the BGG and its variants could definitely be another direction of future research.